# Growth of GaMnAs under near-stoichiometric conditions

V. Avrutin,[a)] D. Humienik, S. Frank, A. Koeder, W. Schoch, W. Limmer, R. Sauer, and A. Waag

Abteilung Halbleiterphysik, Universität Ulm, D-89069 Ulm, Germany

## Abstract

We studied the effect of the V/III flux ratio and substrate temperature on magnetotransport properties and lattice parameters of $Ga_{0.96}Mn_{0.04}As$ grown by molecular beam epitaxy. For all the substrate temperatures, the conductivities and Curie temperatures of the layers were found to increase as the V/III flux ratio approaches 1. The Curie temperature as high as 95 K was achieved for the $Ga_{0.96}Mn_{0.04}As$ samples grown at 240°C and a V/III ratio of about 1.5. The lattice parameter of $Ga_{0.96}Mn_{0.04}As$ increased with decreasing V/III ratio and/or increasing growth temperature. Possible reasons for the effect of V/III ratio on the magnetotransport properties and lattice parameter of GaMnAs are discussed.

[a)] Electronic mail: vitaly.avrutin@physik.uni-ulm.de

In recent years, the ferromagnetic semiconductor (GaMn)As has attracted considerable attention of research community as a promising candidate for spintronic applications[1]. This material is of particular interest due to both its compatibility with existing III-V technology and great recent progress in improving its magnetic properties. (GaMn)As, a supersaturated solid solution of Mn in the GaAs matrix, is fabricated by low-temperature molecular beam epitaxy (LT-MBE). Therefore, the material contains a high density of various defects compensating Mn acceptors. This deteriorates its magnetic properties, because the ferromagnetism in GaMnAs arises from an indirect exchange between magnetic moments of Mn ions mediated by free holes[2]. New insight into physical mechanisms responsible for the compensation in this material and elaboration of efficient techniques for controling its defect structure allowed one to increase the Curie temperature ($T_C$) to 130-170 K.[3-6] To a large extent, this recent progress has been reached due to an understanding of the important role of manganese interstitials, $Mn_I$, as a compensating defect in GaMnAs.[7,8] It was demonstrated that the $Mn_I$ concentration decreases drastically upon post-growth annealing.[7,9] However, other types of compensating point defects such as As antisites, $As_{Ga}$, and As interstitials, $As_I$, as well as their complexes also occur in GaAs-based materials grown by LT-MBE.[10, 11] As reported earlier,[12,13] the concentration of defects in LT-GaAs can be decreased dramatically by growing the material at nearly stoichiometric V/III flux ratios, which improves its optical and transport properties. This finding provides an instrument allowing one to control the defect structure of materials during the epitaxial growth. Indeed, the effect of V/III flux ratio on the lattice parameter of GaMnAs has been observed very recently.[14]

In this paper, we report on the effect of V/III flux ratio and growth temperature on the lattice parameter, magnetotransport, and Curie temperatures in GaMnAs films grown by MBE.

The $Ga_{0.96}Mn_{0.04}As$ films were grown in a solid source Riber 32 MBE system. Epi-ready semi-insulating GaAs(001) substrates were mounted with In on a Mo holder. A valved cracker cell operating in the uncracking mode served as a source of $As_4$. A conventional Knudsen cell and a hot-lip cell were used as sources of Ga and Mn, respectively. The growth rate was about 240 nm/h. First, a 150-nm GaAs buffer layer was grown at a substrate temperature $T_S$ = 580°C, then $T_S$ was lowered during the growth interruption, and the GaMnAs films were grown at three $T_S$ = 190, 225, and 240°C using the same Mo holder. It is well known that the accurate determination of $T_S$ is a severe problem for LT-MBE. However, in order to reveal the effect of the V/III flux ratio on the lattice parameter $a_{GaMnAs}$ of GaMnAs, the accurate control over $T_S$ is essential, because it has a considerable effect on incorporation of excess As and, thus, $a_{GaMnAs}$.[15,16] The value of lattice-parameter expansion, $\Delta a/a$, of LT-GaAs grown under the normal, As-stabilized conditions (typical V/III flux ratio ~3)[17,18] can be used as a measure of an actual growth temperature. [18,19] At each $T_S$ used in this study, we grew reference LT-GaAs layers at a V/III flux ratio of about 3 using the same Mo holder as for the growth of GaMnAs films, and the growth temperatures were calibrated with the use of the $\Delta a/a(T_S)$ dependence.[18] The composition of the GaMnAs films was estimated from flux measurements, and the high reproducibility of the Mn content from sample to sample was confirmed by x-ray photoelectron spectroscopy (XPS).

To determine the minimum V/III flux ratios at which the growth of single-crystal $Ga_{0.96}Mn_{0.04}As$ was possible, we grew a series of samples at various $T_S$ and V/III ratios. The growth time for each sample was 1 h. Reflection high-energy electron diffraction (RHEED) monitoring revealed that, when the V/III ratio was below a certain (threshold) value, clusters appeared on the surface which led to polycrystalline growth. The onset of clustering was featured by the change from a streaky (1x2) RHEED pattern to spotty one. As shown in Fig. 1, the threshold flux ratio is a strong function of $T_S$. At 190°C, the growth is possible at V/III ratio of about 1, *i.e.*, very close to the stoichiometric conditions. It is worth noting that, at a given $T_S$, the growth time it takes for clustering to begin decreases with decreasing the V/III ratio.

To study the influence of V/III flux ratio and $T_S$ on the lattice parameter of GaMnAs and its magnetotransport properties, we grew 200-nm-thick $Ga_{0.96}Mn_{0.04}As$ layers using high or low V/III flux ratios at 190, 225, and 240°C. At each $T_S$, the flux ratios corresponding to the data points shown in Fig. 1 are referred to as low ones (near-stoichiometric conditions). By the high V/III value, we mean the ratio of about 3 providing the As-stabilized conditions for the growth of high-temperature GaAs. The thickness of GaMnAs layers was estimated from the RHEED oscillations and x-ray pendeloesung fringes.

The effect of substrate temperature and V/III flux ratio on the lattice parameter of GaMnAs layers was studied by high resolution x-ray diffraction (HRXRD) using a Siemens D5000 diffractometer. The (004) diffraction patterns recorded from all the samples showed symmetric and narrow GaMnAs peaks with strong pendeloesung fringes, indicating the absence of nonuniformities in the lattice parameter. Figure 2 summarizes the effects of the V/III ratio and $T_S$ on the lattice parameter of $Ga_{0.96}Mn_{0.04}As$. As seen from Fig. 2, $a_{GaMnAs}$ rises with increasing $T_S$ and/or decreasing flux ratio. As the growth temperature rises, the difference between $a_{GaMnAs}$ in the layers grown at different flux ratios decreases and virtually vanishes at 240°C. It should be mentioned that similar increase in $a_{GaMnAs}$ with decreasing the V/III ratio has been reported by Sadowski and Domagala[14] for the layers grown using $As_2$ species at 230°C. At the first glance, the increase in $a_{GaMnAs}$ with increasing $T_S$ runs counter to the expectations based on the well-known fact that the lattice expansion of LT-GaAs due to the incorporation of excess As decreases with increasing growth temperature.[17] However, measuring dependences of $a_{GaMnAs}$ on the Mn content $x$ for the layers grown at two different $T_S$ = 200 and 240°C, we found that the curve constructed for higher $T_S$ starts from lower lattice constant of LT-GaAs and runs steeper intersecting with the curve plotted for lower $T_S$ at about 2.5% of Mn (see the inset in Fig. 2). The layers grown at higher $T_S$ show higher lattice parameters in the concentration range above 2.5%-3% of Mn. Similar behavior was observed also by Schott *et al.*[15] for $T_S$ = 220 and 270°C. The above results suggest that the contributions from all defect species ($Mn_{Ga}$, $Mn_I$, $As_{Ga}$, $As_I$) and their complexes to the lattice expansion should be taken into account. Unfortunately, the effects of different defects on the lattice parameter of GaMnAs cannot be separated based on the available experimental data and demand particular study.

The transport properties of GaMnAs layers were studied by resistivity and Hall measurements on photolithographically defined Hall bars in the temperature range from 4.2 to 300 K in magnetic fields up to 14 T. The Curie temperatures were determined from the temperature dependences of remanent part of Hall resistance, $R_{xy}(0)$, divided by sheet resistance, $R_{xx}(0)$, measured at zero magnetic field after the sample had been magnetized in a magnetic field of about 2 T. The $R_{xy}(0)/R_{xx}(0)$ value is proportional to the remanent magnetization $M$ of the material. The details of this method were described earlier by Ohno *et al.*[20] The V/III flux ratio was found to have a strong effect on the conductivity of the GaMnAs samples. For the samples grown at 190°C, the decrease in V/III ratio gives rise to the dielectric-metal transition. The metallic samples grown at 190°C show $T_C$ ~ 35 K, whereas the Curie temperature for the nonmetallic samples cannot be determined from the transport measurements. All layers prepared at higher $T_S$ are metallic, and those grown at lower V/III ratios show higher conductivities and higher $T_C$ = 70 and 95 K for $T_S$ = 225 and 240°C, respectively, whereas the Curie temperatures in counterparts prepared using high V/III ratios are only 45 and 60 K. Figure 3 shows the temperature dependences of $R_{xy}(0)/R_{xx}(0)$ for the samples grown at $T_S$ = 240°C at different flux ratios. One can see that both the $R_{xy}(0)/R_{xx}(0)$ curves deviate from the Curie-Weiss behavior. This can be attributed to inhomogeneities of magnetic properties[21] caused by nonuniform distribution of Mn magnetic moments and/or free holes. The later is more probable, because the HRXRD data point to the uniformity of the lattice parameter. Our earlier electrochemical capacitance-voltage (ECV) measurements carried out on a variety of GaMnAs samples[22] revealed the gradient of free-carrier concentration across the film thickness with the highest hole density in the near-surface region. Based on this finding, we suggest that the precisely the topmost layer determines the Curie temperature. It is important to note that the $R_{xy}(0)/R_{xx}(0)$ curve measured for the sample grown at low V/III ratio shows much more complex behavior, indicating more pronounced gradient in hole concentration.

The hole concentration $p$ in the near-surface region of GaMnAs layers was assessed from the shape and spectral position of a Raman line corresponding to the coupled mode of longitudinal optical (LO) phonon and hole plasmon. With increasing $p$, this line drastically broadens and shifts from the frequency of the LO phonon to that of the transverse optical (TO) phonon. In more detail, this technique was reported elsewhere.[23] It is important to mention that, for the excitation wavelength of 514 nm used in our study, the probing depth of Raman spectroscopy is only about 50 nm. Thus, according to our previous ECV data,[22] the Raman spectra reflect the highest $p$ in our samples. Comparing the Raman spectra (Fig. 4), we can deduce that, for all the growth temperatures, the hole densities are considerably higher in the samples grown at low V/III ratios, in a good agreement with the $T_C$ values. Note that the spectra of the samples grown at 190°C differ dramatically. In line with the transport data revealing nonmetallic behavior of conductivity in the sample grown at the high V/III ratio, the coupled mode in this sample appears as a narrow line near the LO-phonon frequency, whereas it is broadened drastically and shifted towards the TO-phonon frequency for low V/III ratio. The samples grown at 225 and 240°C using low flux ratios show the substantial difference in Curie temperatures (70 and 95 K, respectively), although their Raman spectra are almost identical. According to the model proposed by Dietl *et al.*,[2] this difference can be attributed to the difference in hole density and/or Mn-spin concentration $x$: $T_C \sim x \times p^{1/3}$. However, our estimates of $p$ from the Raman line shape analysis show that the difference in $p$ between these two samples cannot be sufficiently large to provide the observed difference in $T_C$. Therefore, the improvement of the Curie temperature is more likely due to higher concentration of Mn spins (*i.e.*, Mn ions on Ga sites) in the near-surface region of the sample grown at 240°C.

Based on the obtained results, we can suggest that a competition between Mn and excess As for Ga lattice sites takes place at the surface of a growing GaMnAs layer. The amount of excess As should decrease with decreasing the V/III ratio and/or increasing $T_S$. Therefore, at low V/III ratios and/or high $T_S$, Mn is more likely incorporated into a lattice position, where it acts as an acceptor, than into an interstitial site, where it behaves as a compensating donor and can form antiferromagnetically ordered $Mn_I$-$Mn_{Ga}$ pairs by occupying positions adjacent to $Mn_{Ga}$.[8] The decrease in the amount of excess As also leads to lower acceptor compensation. As a result, the Curie temperature increases in the samples grown at low flux ratios. However, further investigations of depth distributions of defects and free carriers are necessary to clarify the complex behavior of magnetization for the films grown at low V/III ratios.

In summary, we have demonstrated that the V/III flux ratio is an important instrument allowing one to control the magnetotransport properties of GaMnAs layers during the MBE growth. For all the substrate temperatures, the hole densities, conductivities, and Curie temperatures of the layers grown at low V/III ratios was demonstrated to be higher than those of the films grown at high flux ratios. The Curie temperature as high as 95 K was achieved for the samples grown at a V/III ratio of about 1.5 and $T_S$ = 240°C. We also found that $a_{GaMnAs}$ increases with decreasing the V/III ratio and/or increasing the growth temperature.

We thank H. Rauscher for the XPS measurements of GaMnAs layers. This work was supported by the Deutsche Forschungsgemeinschaft, DFG Wa 860/4-1.

FIGURE CAPTIONS

Fig. 1. Flux ratio *vs.* substrate temperature diagram. The curve represents the threshold V/III flux ratio for the 1-hour growth of $Ga_{0.96}Mn_{0.04}As$ layers.

Fig. 2. Lattice parameter of $Ga_{0.96}Mn_{0.04}As$ as a function of substrate temperature for the samples grown at low and high V/III flux ratios. The lines are just to guide the eye. The inset shows the lattice parameter of GaMnAs layers grown at 200°C (solid squares) and 240°C (open circles) as a function of Mn content $x$.

Fig. 3. Temperature dependences of remanent part of $R_{xy}(0)$ divided by zero-field resistivity $R_{xx}(0)$ measured for $Ga_{0.96}Mn_{0.04}As$ samples grown at 240°C at low and high V/III flux ratios.

Fig. 4. Raman spectra recorded from $Ga_{0.96}Mn_{0.04}As$ layers grown at low and high V/III flux ratios and various substrate temperatures. The dashed lines mark the TO- and LO-phonon frequencies for the GaAs substrate.

Fig. 1., Avrutin et al., Appl. Phys. Lett.

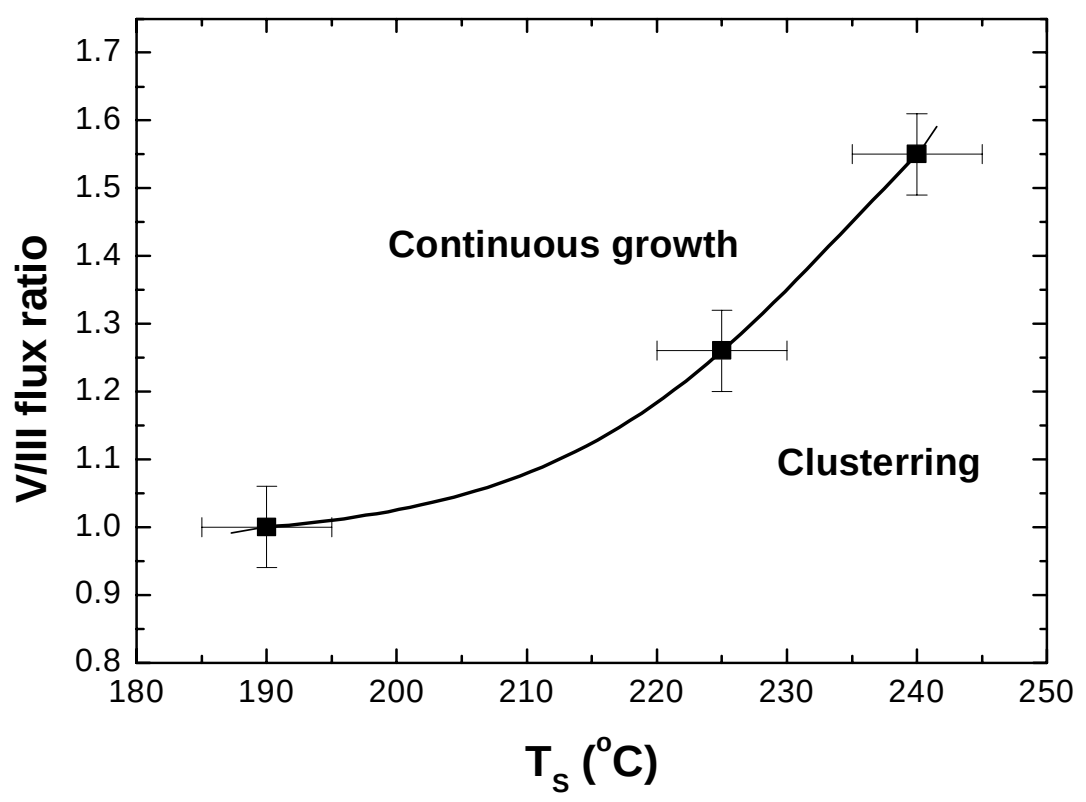

Fig. 2., Avrutin et al., Appl. Phys. Lett.

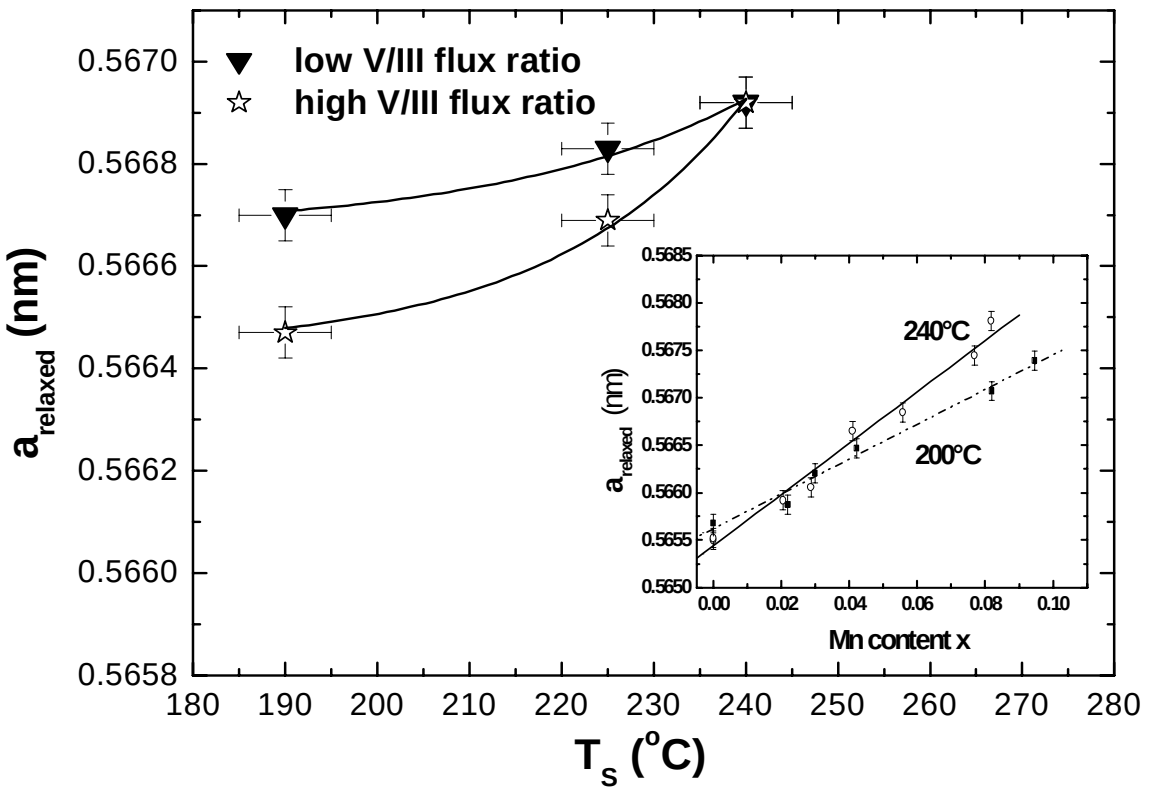

Fig. 3., Avrutin et al., Appl. Phys. Lett.

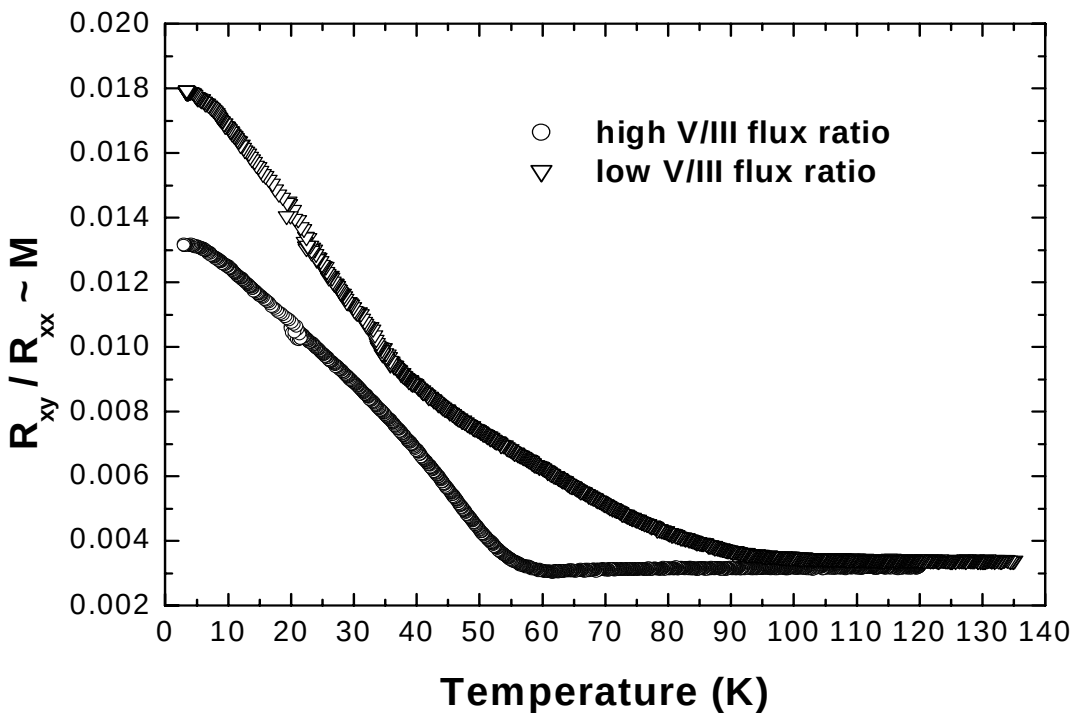

Fig. 4., Avrutin et al., Appl. Phys. Lett.

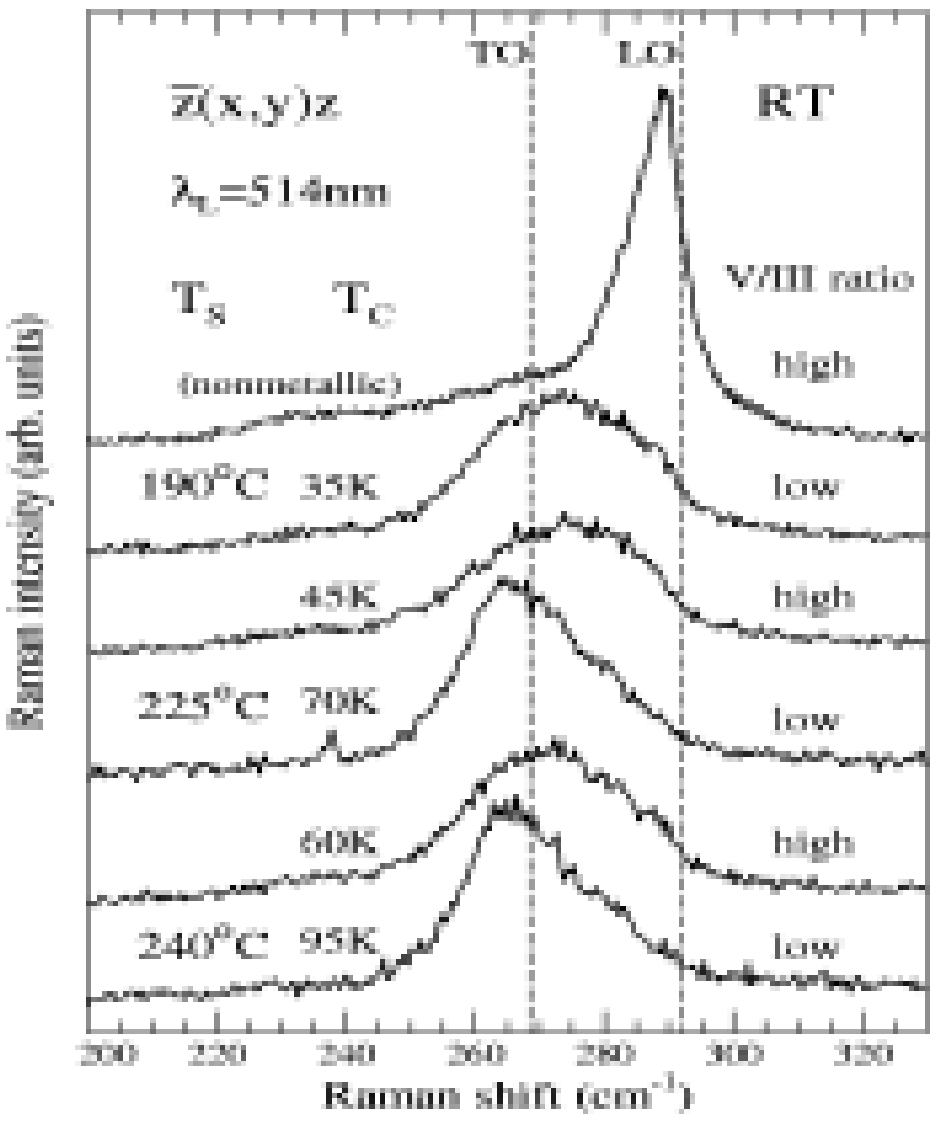